# Hilbert entropy for measuring the complexity of high-dimensional systems

Seong-Gyun Im [1,†], Taewoo Kang [1,†], and S. Joon Kwon [1,2,3,4]*

[1]School of Chemical Engineering, Sungkyunkwan University (SKKU), Republic of Korea

[2]SKKU Institute of Energy Science & Technology (SIEST), SKKU, Republic of Korea

[3]Department of Semiconductor Convergence Engineering, SKKU, Republic of Korea

[4]Department of Future Energy Engineering, SKKU, Republic of Korea

\* To whom correspondence should be addressed: sjoonkwon@skku.edu

[†] These authors contributed equally to this work.

**Abstract**

Measuring the complexity of high-dimensional data in physical systems becomes a critical factor in determining the information and quality of the systems. However, traditional metrics, such as Lyapunov exponent, fractal dimension, and information entropy, are limited in measuring contextual higher-dimensional data in that they do not elucidate the intrinsic nature of physical systems. Herein, we introduce a novel methodology for quantifying the complexity of high-dimensional data through dimension reduction yet retaining context using a space-filling curve such as the Hilbert curve along with generalized entropy measures. We validate this methodology in measuring critical phenomena, including phase transitions in spin and percolation models. Our findings demonstrate a high degree of concordance between the Hilbert entropy and theoretical phase transition points. Moreover, we further proceed to an exploration of the hidden relationship between the Hilbert entropy and the fractal dimension, such as a linear relationship between scaling exponent and the Euclidean dimension of scale-invariant 2D/3D geometries. The present methodology offers a promising new framework for understanding and analyzing complex systems in higher dimensions, with potential applications across various fields of physics.

The complexities of dynamic systems have been thoroughly studied in statistical physics and nonlinear dynamics[1–4]. In particular, it was proposed that the complex nature of low-dimensional systems, such as time series data, with chaotic characteristics could be revealed by several indices such as Lyapunov exponent[5,6], fractal dimension[7,8], and variants of Renyi entropies[9–14]. These conventional metrics measuring complexities in dynamic systems can quantify part of the system characteristics; however, they are limited in revealing the hidden nature of higher dimensions than a 1D system. There have been several attempts to expand existing 1D complexity metrics to 2D[15–18]. For instance, the graph-isomorphism-based approach can be used to measure the topological complexity of granular systems[19,20]. It should also be noted that the expansion has not been tried to 3D or higher dimensions. However, simply extending the complexity metrics to higher-dimensional contexts, especially those encompassing two or higher dimensions, presents significant challenges, such as numerical artifacts and loss of local information. These challenges make the simple extension inappropriate for measuring the complexity of high-dimensional data. To address this problem, reducing high-dimensional data to a 1D array followed by applying the 1D complexity metrics would be desirable. Unfortunately, the simple conversion of high-dimensional data into a 1D vector is irrelevant in correctly addressing the locality of the original data[21]. To address the locality problem, the space-filling curve (SFC)-based data mapping can be proposed. The SFC (or Peano curve), proposed by Peano in 1890[22], is a curve that continuously visits every point only once in discrete lattice points or cells. As its name suggests, SFC allows mathematical equivalence of the points array connected in a curve concerning the original higher dimensional data, keeping locality. Among well-known SFCs, the Hilbert curve[23] is known to preserve locality best in most cases compared to other SFCs due to attaining the local connectivity of the nearest neighbors in the lattices in higher dimensions[21,24]. Therefore, the SFC-based method allows a mathematically relevant way of dimension reduction. Onto the reduced dimension, the complexity can be measured by entropy considering the intrinsic nature of the data, such as binary, continuous, discrete, logic, and so on.

**Dimension reduction based on space-filling curve**

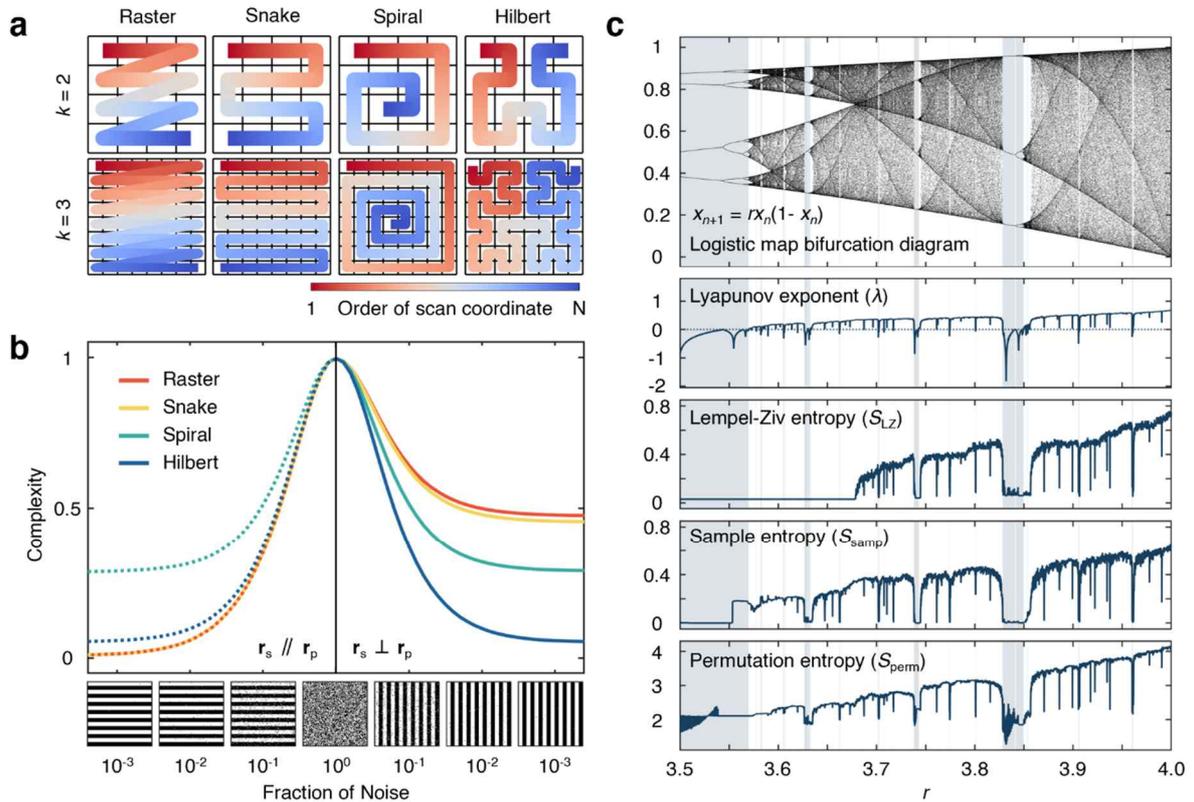

**Fig. 1: Hilbert curve and complexity measures. a**, Scanning methods (raster, snake, spiral, and Hilbert curve, from right to left) for flattening the 2D data into a 1D vector. The red dot represents the randomly chosen two points on the lattice. The degree of distance uncertainties with respect to the recursion order $k$ for mapping functions is also shown at the bottom. **b**, Complexity measured as a function of noise fraction for various scan methods. The dotted line and solid line denote the scanning direction ($\mathbf{r}_s$) parallel or perpendicular to the direction of the stripe patterns ($\mathbf{r}_p$), respectively. **c**, Normalized values of the entropies as a function of the Lyapunov exponent $\lambda$ distinguishing the chaotic and non-chaotic behavior of the logistic map.

Reliable scanning methodologies are required to reduce high-dimensional data into a 1D vector, preserving locality nature. Basically, it should be noted that it is impossible to completely preserve all the local properties of the high-dimensional data when converting it to lower-dimension. In addition, every dimensional reduction is associated with the introduction of unnatural correlation of long-distance data points, which distorts the overall data structures as well as information distribution. For instance, the row- or column-wise raster scan (Fig. 1a) induces unexpected discontinuity at the system's boundaries, which overestimates the system's complexity or introduces boundary effects. The row-wise snake scan keeps the row-wise information (i.e., neighbors) while inducing loss in column-wise information (Fig. 1b). For the spiral scan, information close to the center of the scan can be preserved; however, the more the information is far from the center, the higher the probability of over or underestimation in a specific direction. As the system size expands, the scanning bias exacerbates, leading to a corresponding increase in uncertainty and a decrease in measure reliability (Supplementary Fig. 1).

To address the bias problem, we devised a mapping method other than existing manners. The SFC can be generated in a recursive manner for any space in any shape or form. In particular, the Hilbert curve is advantageous in space-filling of high-dimensional spaces and a variety of discrete data with different shapes, including anisotropic or spherical lattices (Supplementary Fig. 2). For instance, the $k$th order Hilbert curve in $n$ dimension can completely fill discrete lattices with cells of $\prod_{i=1}^{n} 2^k$. To test the performance of the Hilbert curve-based scanning method, we compared the complexity (permutation entropy, which will be elaborated upon in further detail subsequently) as a function of noise fraction for different scanning methods. The salt-and-pepper noise was added to the horizontal and vertical stripe patterns with a fraction ranging from $10^{-4}$ to 1. Fig. 1b shows that the Hilbert curve-based flattening yields negligible disparities between the vertical and horizontal stripe patterns in terms of complexity. It should also be noted that the Hilbert curve showed a marginal bias from 0, an ideal entropy value for the simplest pattern (i.e., without noise), for the pattern with minimal noise. In contrast, other scan methods exhibited considerable variations in complexity depending on the scanning directions (raster and snake scans) or substantial bias (spiral scan).

Next, we searched for relevant complexity measuring methods for the dimension-reduced data (the flattening method is shown in Supplementary Fig. 3). For this, we tested a representative dynamic system, a logistic map, which generates a well-known self-similar bifurcation pattern as a function of a parameter $r$ in a recursion function: $x_{n+1} = rx_n(1-x_n)$ for $x_n \in [0,1]$. For the logistic map (Fig. 1c), the Lyapunov exponent ($\lambda$) can work as a reference for the universal complexity measure. Concerning $\lambda$, we tested several entropies such as Lempel-Ziv entropy ($S_{LZ}$)[25], sample entropy ($S_{samp}$)[26], and permutation entropy ($S_{perm}$)[27]. As compared in Fig. 1c, each entropy captures most of the critical bifurcation points in the map. However, each entropy is also limited in detecting every single bifurcation point when compared to $\lambda$ (Supplementary Fig. 4). Unfortunately, it should also be noted that most high-dimensional data with complex structures are not always associated with the governing function like logistic map, and therefore, $\lambda$ is not necessarily always available for the measurement of the complexity. Therefore, rather $\lambda$, the entropy-based measures should be carefully employed in a complementary manner by combining or selecting different entropies for specific data types. By combining the Hilbert curve and entropy measures, we can use, namely, the Hilbert entropy to measure the complexity of high-dimensional data as precisely as possible.

**Spin models**

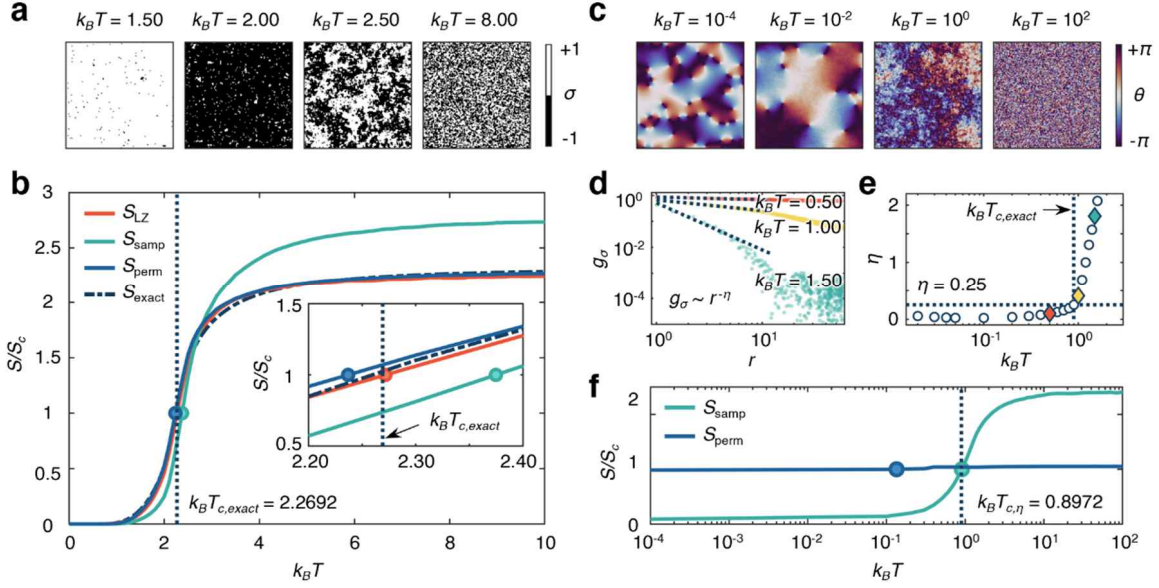

**Fig. 2: Spin models. a**, Snapshots of 2D square lattices of Ising model at different temperatures. **b**, The entropy increase with respect to the temperature. Inset plot shows an expanded region of the plot showing the inflection point detected by different entropy measures. Filled circles and the vertical dotted line are shown to indicate the inflection points. **c**, Snapshots of the 2D XY model at different temperatures. **d** Power-law decaying behavior of orientational correlation function, $g_\sigma(r)$, at $k_BT$ = 0.50, 1.00, and 1.50 (**d**). The dotted slope lines in **d** denote the envelope functions of $g_\sigma(r)$ for each temperature. **e**, the evolution of the power exponent $\eta$ with increasing the temperature. The vertical line in **e** denotes the point over which the orientational order diminishes ($\eta \geq 0.25$). **f**, Normalized entropies as a function of the temperature. Filled circle makers represent the inflection points and a vertical dotted line is the critical temperature for the orientational order where $\eta = 0.25$ ($k_BT = 0.8972$).

To validate the Hilbert entropy, we first tested well-known physical models based on the 2D spin models. We used a 2D Monte Carlo (MC) method (dimension of $2^7 \times 2^7$ square lattice) for the simulation of 2D spin models. The validity can be confirmed by testing the Hilbert entropy's detection of the critical points of the order-to-disorder transition of the models. Fig. 2a shows the simulated 2D Ising model with discrete spin states ($\sigma$) of -1 (spin down) and 1 (spin up). We measured the system's complexity using Hilbert entropy of the dimension-reduced vector containing binary spin arrangement in each simulated lattice at different temperatures and compared it with the theoretical entropy of the Ising model ($S_{exact}$)[28]. As shown in Fig. 2b, three entropies ($S_{LZ}$, $S_{samp}$, and $S_{perm}$) can qualitatively track the order-to-disorder transition of the 2D Ising model with increasing temperature. However, $S_{samp}$ overestimates the complexity for higher temperatures, while $S_{LZ}$ and $S_{perm}$ correctly track the entropy evolution and precisely detect the critical point ($T_C$). For instance, $k_B T_C$ derived from the $S_{LZ}$, and $S_{perm}$ are 2.2717 and 2.2370, respectively, which are close to the theoretically exact value of 2.2692[28]. In particular, compared to $S_{samp}$, $S_{LZ}$ is advantageous in measuring the intrinsic complexity in binary system due to the fact that $S_{LZ}$ considers every possible combinations of binary signals, while $S_{samp}$ does not consider the local order of the signal arrangement. The lack of detailed consideration of the local order leads to a decrease in the entropy value at the $k_B T_C$, which gives rise to overestimated normalized entropy for the $S_{samp}$.

Unlike the 2D Ising model with discrete spins, the 2D XY model considers continuous spin orientation ($\theta$) ranging from $-\pi$ to $\pi$, and therefore, $S_{samp}$ and $S_{perm}$ are relevant in measuring the complexity of the disordered system. We used the MC simulation for the 2D XY model to obtain simulated samples (Fig. 2c). Considering the fact that the 2D XY model undergoes a Kosterlitz-Thouless (KT) phase transition, we tracked the orientational correlation ($g_\sigma(r)$), from which the power-law dependence of $g_\sigma(r)$ with respect to the distance vector $r$ can be analyzed with the decay exponent ($\eta$) of the correlation (Fig. 2d). According to the Kosterlitz-Thouless-Halperin-Nelson-Young (KTHNY) model, the exponent $\eta$ has a threshold value of 1/4, above which the system evolved into a disordered system by losing orientational order[29]. By tracking the value of $\eta$ as a function of the temperature, it is possible to extract the critical temperature such that $k_B T_C = 0.8972$, as shown in Fig. 2e. Interestingly, we observed that the detected critical temperature from the Hilbert entropy with $S_{samp}$, $k_B T_C = 0.9078$, is close to the value extracted from the MC simulation (Fig. 2f). In contrast, $S_{perm}$ did not exhibit reliability in measuring the critical temperature and overestimated the system's complexity at low temperatures. This discrepancy is due mainly to the intrinsic nature of the measurement of $S_{perm}$ in which every possible ordinal pattern in a given embedding dimension is counted. Then, $S_{perm}$ detects unnecessary detailed micro spin distribution patterns in very low temperatures, which are prone to erroneous detection of the ghost critical point(s) ($k_B T \sim 10^{-1}$). From the validity test, we can conclude that Hilbert entropy with $S_{samp}$ for continuous or binary data and with $S_{LZ}$ and $S_{perm}$ for discrete data can be generally applied to physical systems suffering critical phenomena.

**Percolation model**

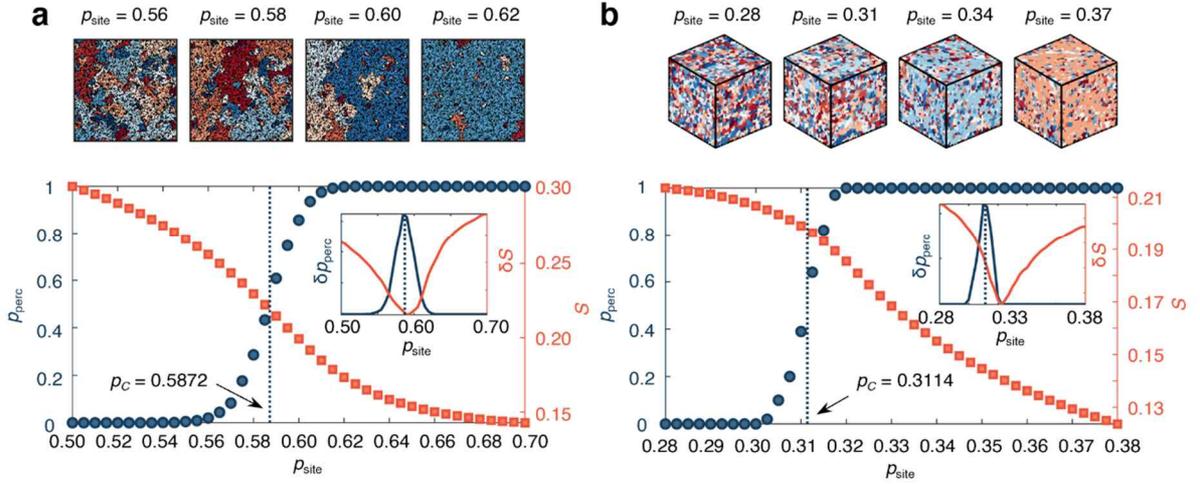

**Fig 3: Percolation models. a, b**, Site percolation model for 2D (**a**) and 3D (**b**) systems. Top figure panels show the distribution of clusters with the different values of the site probability, $p_{site}$. The bottom plots show the behavior of the percolation probability, $p_{perc}$ (navy-colored filled circles), and the Hilbert entropy measured with $S_{LZ}$ (red-colored filled squares) are shown as a function of $p_{site}$ (inset: first derivative of $p_{perc}$ and $S_{LZ}$ with respect to $p_{site}$, denoted as $\delta p_{perc}$ and $\delta S$, respectively). The vertical dotted lines indicate the percolation threshold, $p_C$, which is the inflection point of the $p_{perc}$.

Next, we extended the study of the application of Hilbert entropy in higher dimensional and different physical systems. For this, we employed 2D and 3D percolation models in discrete lattices (site percolation). Different from the 2D Ising or XY model, the percolation model tracks the structural correlation of the percolation network by estimating the percolation probability $p_{perc}$. In the site percolation model, by incrementing the probability (i.e., site probability; $p_{site}$) of the lattice site being occupied, the percolation threshold ($p_C$) at which the likelihood of the entire system undergoing percolation surges can be calculated. When $p_{site} < p_C$, small clusters dominate the system, while the probability of clusters spanning the entire system abruptly increases when $p_{site} > p_C$. As such, the transition from multiple small, disconnected clusters to a single, large, connected cluster exemplifies the second-order phase transition. Therefore, 2D or 3D percolation lattice models are relevant to test the applicability and validity of the Hilbert entropy in higher dimensional data. For the modeling, we used MC simulation to generate the 2D or 3D lattices (dimensions of $2^7 \times 2^7$ for 2D and $2^6 \times 2^6 \times 2^6$ for 3D) with different values of $p_{site}$ (Figs. 3a and 3b). For each of the lattices, we calculated $p_{perc}$ as a function of $p_{site}$. In addition, we also calculated the complexity of the lattices using Hilbert entropy along with $S_{LZ}$, considering the discrete and non-ordinal nature of the lattices. To detect the $p_C$, we calculated the first derivative of the entropy. As compared in Fig. 3a, we observed that the Hilbert entropy can detect the $p_C$ in 2D percolation lattices with slight discrepancy ($p_C$ from $S_{LZ}$ = 0.5916; $p_C$ from MC simulation = 0.5872). It should be noted that the Hilbert-entropy-based $p_C$ is close to the reported values, such as 0.5927[30–32]. We extended the analysis for 3D lattices (Fig. 3b) and observed that the Hilbert entropy is also capable of identifying $p_C$ with minimal deviation from the reference ($p_C$ from $S_{LZ}$ = 0.3225; $p_C$ from MC simulation 0.3114), which is shown to be still effective when compared to the theoretical value for 3D percolation such that $p_C$ = 0.3116[32,33]. The discrepancies are believed to be due to the simulation box effect. From the analysis of 2D and 3D percolation lattices, we can conclude that Hilbert entropy with a relevant combination of information entropy ($S_{LZ}$) is capable of measuring the intrinsic complexity of high-dimensional data.

**Measurement of intrinsic complexity of scale-invariant systems**

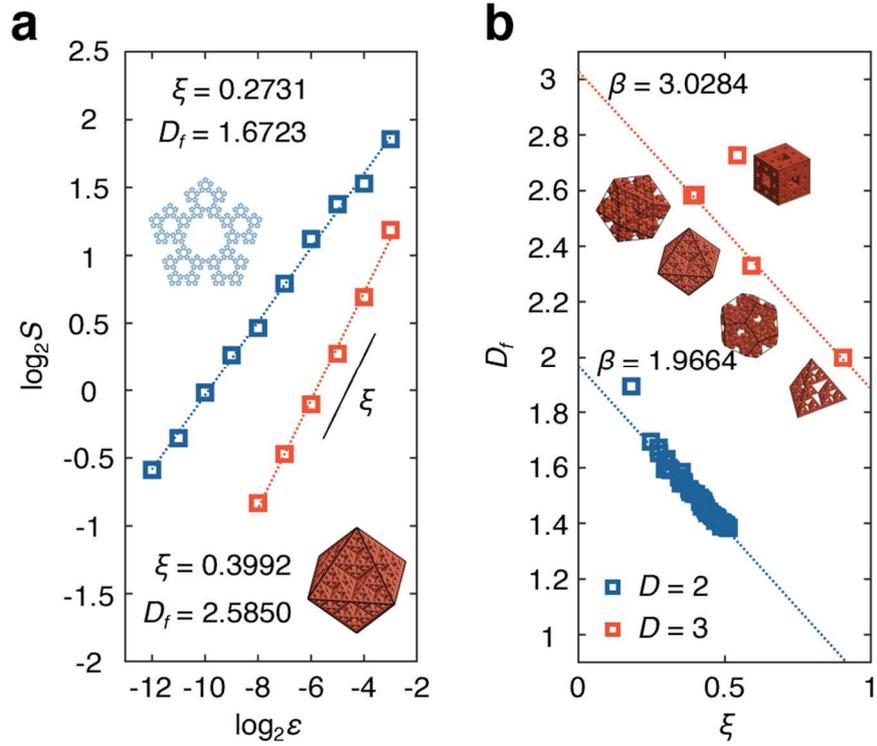

**Fig. 4: Relationship between the Hilbert entropy and the fractal nature of scale-invariant systems.**
**a**, Power law dependencies of the Hilbert entropy (with $S_{\text{perm}}$) as a function of a box size $\varepsilon$ (relative to the original size of the circular or spherical container encompassing the fractal geometry)), in 2D (number of edges ($E = 5$), red open squares) and 3D (number of faces ($F = 8$), blue open squares) iterated function system (IFS)-based fractals (with the fractal dimension of $D_f$). The linearities were analyzed with the slope (power exponent $\xi$) of the linear regression for each IFS. **b**, The linear relationship of the power exponent $\xi$ and the fractal dimension $D_f$ for 2D (red open squares) and 3D (blue open squares) IFSs. The dotted line indicates the regression line. For 3D IFSs, corresponding IFSs are presented next to the corresponding data points.

One of the notable representative properties of the dynamic systems suffering continuous phase transition is the scale invariance accompanied by self-similarity. For the quantitative measurement of self-similarity, the fractal dimension ($D_f$) can be used. To test and find a hidden correlation between the fractal properties and the entropy, we calculated the Hilbert entropy of a variety of iterated function systems (IFSs) with different numbers of edges ($E$) or faces ($F$) such as penta-flake ($E = 5$), or octahedron IFS fractal ($F = 8$), as shown in Fig. 4a. In principle, the Hilbert curve is compatible with fractal geometries generated by different recursion numbers, which can correspond to the order of the Hilbert curve. Therefore, it is expected to find a strong correlation between $D_f$ and the entropy. It should be noted that conventional information entropy does not consider the self-similar and recursive nature of the fractal geometries.

We first investigated a relationship between the resolution of the fractal geometries and the entropy. To quantify the resolution, we used different box sizes $\varepsilon$. The exemplary quantization process with respect to $\varepsilon$ is shown in Supplementary Fig. 5. The entropy should increase with increasing the value of $\varepsilon$. In other words, $\varepsilon$ is inversely proportional to the order of the Hilbert curve. Specifically, systems with a high $D_f$ experience a gradual reduction in the entropy with increasing $\varepsilon$, as the neighboring boxes have a mitigating effect on the entropy reduction. Conversely, in systems with low $D_f$, the influence of neighboring boxes is minimal, leading to a more pronounced decrease in entropy. Due to the self-similarity, clear power-law dependences between the $S_{\text{perm}}$-based entropy and $\varepsilon$ are expected.

Interestingly, as shown in Fig. 4a, the Hilbert entropy exhibited a power law with a particular exponent $\xi$ corresponding to the particular IFSs, which can be defined as

$$\xi = -\lim_{\varepsilon \to 0} \frac{\log S(\varepsilon)}{\log(1/\varepsilon)}. \tag{1}$$

As described in equation (1), $\xi$ can be considered as an indicator to infer $D_f$ for various self-similar but complicated geometries. It means that $\xi$ can be used to uncover the fractal nature of complex high-dimensional systems. To test this conjecture, we analyzed the relationship between $D_f$ and $\xi$ of the given 2D and 3D IFSs. As shown in Fig. 4b, we observed clear linearities between $\xi$ and $D_f$. Remarkably, the linearities commonly reveal a simple relationship of $D_f$ and $\xi$ such that $D_f = -\alpha\xi + \beta$. The values of the coefficients are $\alpha = $ 1.1637 and 1.1453, while the constant values of $\beta$ are 1.9664 and 3.0284 for 2D and 3D IFS, respectively. The values of $\xi$ and $D_f$ for other IFSs are shown in Supplementary Fig. 6. This implies that there can be an ideal and simple linearity between the real $D_f$ and the scaling exponent of the entropy such that

$$D_f = -\xi + D, \tag{2}$$

where $D$ denotes the Euclidean dimension. To elucidate the mathematical principle underlying this linearity, we can refer to the exemplary 1D fractional Brownian motion (fBm), $B(r)$ at position $r$ satisfying the autocorrelation such that

$$\langle B(r+\varepsilon)B(r)\rangle = \frac{V_H}{2}\left(r^{2H} + |r+\varepsilon|^{2H} - |\varepsilon|^{2H}\right), \quad V_H = \Gamma(1-2H)\frac{\cos(\pi H)}{\pi H} \tag{3}$$

Where $\Gamma(\cdot)$ denotes the gamma function and $H$ is for Hurst exponent of the 1D vector. The 1D fBm can be considered as the dimension-reduced higher dimensional data based on Hilbert curve. To calculate the entropy of 1D fBm, we can calculate the autocorrelation of the difference of $B(r)$ such that, $\langle C(r+\varepsilon)C(r)\rangle$, where $C(r) = B(r+1) - B(r)$. It can be shown that

$$\langle C(r+\varepsilon)C(r)\rangle = \frac{V_H|\varepsilon|^{2H}}{2}\left(\left(1+\frac{1}{\varepsilon}\right)^{2H}+\left(1-\frac{1}{\varepsilon}\right)^{2H}-2\right)$$

$$\approx \frac{V_H|\varepsilon|^{2H}}{2}\left(4H(2H-1)\varepsilon^{-2}\right) \quad (4)$$

$$= 2H(2H-1)V_H\varepsilon^{2H-2} \sim \varepsilon^{2(H-1)}.$$

The autocorrelation function value should be proportional to the squared value of the entropy. It should also be noted that the entropy value should be approaching zero when $\varepsilon \to 0$. Then, from equation (4), we can find that

$$S \sim \varepsilon^{H-1}. \quad (5)$$

From equation (5), we can find that $\xi = H-1$. For 1D fBm generated by self-similar sequence given $H$, the fractal dimension, $D_f$, satisfies a relationship of $D_f = D - H + 1$ [34], which leads to an ideal linearity we found above such that $D_f = -\xi + D$.

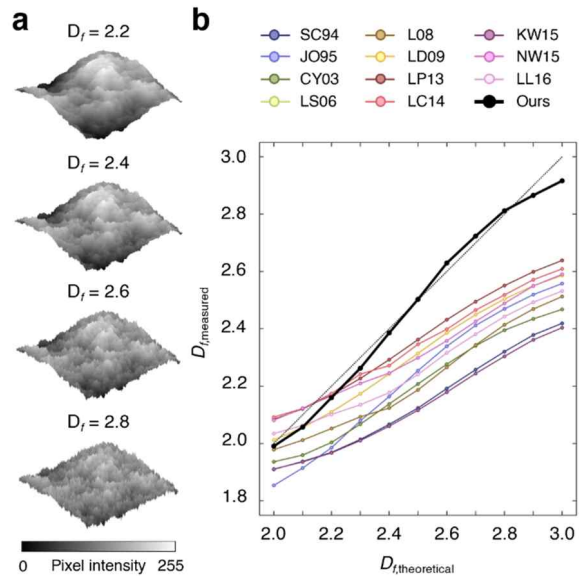

**Fig. 5: Hilbert entropy to measure the fractal dimension of the synthetic high-dimensional data.**

**a**, Fractional Brownian surface (fBs) with different fractal dimensions ($D_f = 2.2, 2.4, 2.6$, and $2.8$, from top to bottom) generated by controlling the Hurst exponent ($H$). The z-axis indicates the pixel value for the grayscale image (0 to 255). **b**, Comparison between the power exponent $\xi$-based fractal dimension (black) and box-counting method-based fractal dimensions (colored) retrieved from ref.[35]. The initial of authors and publication date are presented on the legend. The black dotted line indicates the ideal fractal dimension calculated from the relationship with the Hurst exponent: $D_f = D - H + 1$. The closer the curve is to the dotted line, the more accurate the method is.

Next, we further explore and apply the observed universality of the power exponent $\xi$ as a complexity metric for general high-dimensional complex data, which is not limited to data with binary properties. Although the ideal linearity of $D_f = D - \xi$ is proven its applicability to the binary self-similar system, the more generalized linearity such that $D_f = -\alpha\xi + \beta$ can provide unique information defined by a combination of $\alpha$ and $\beta$ for higher dimensional systems exhibiting properties other than binary. As revealed in equation (2), this finding suggests that the intrinsic dimension of the high-dimensional data can be inferred even for the non-binary systems which have been limited in determining the $D_f$. For example, the conventional box-counting methods are known to be unable to evaluate $D_f$ of the grayscale image reliably[35]. This is mainly because of the fact that the grayscale image is not associated with a dimension on a purely mathematical basis. Then, this leads to an increasing discrepancy between the calculated fractal dimension and the ideal fractal dimension in the case of 2D fBm when the fractal dimension approaches $3$[35–37]. Instead of conventional box-counting methods, therefore, the present Hilbert entropy-based approach can be advantageous in obtaining the more realistic and exact value of the fractal dimension of general complex high-dimensional data, including grayscale images. To test the performance of the Hilbert entropy-based measurement of the fractal dimension, we tested fBm-based images (i.e., fractional Brownian surface (fBs) images) with theoretically defined fractal dimensions (Fig. 5a). For the calculation of the $\xi$ in grayscale image having pixel values from 0 to 255, we used $S_{\text{samp}}$ which has proven its performance in 2D XY model. The dimension of the grayscale image is should be equal to 3 (i.e., Euclidean dimension of 2 in planar dimension with an additional single degree of freedom corresponding to the pixel value distribution, considering the image as a 3D surface with the height of brightness). As shown in Fig. 5b, the inferred fractal dimension from the linearity of $D_f = -\alpha\xi + \beta$ works better than the conventional box-counting methods. In particular, $S_{\text{samp}}$-based $D_f$ measure exhibits reduced discrepancy even in the case of the theoretical fractal dimension approaching 3. This shows that the present method based on the Hilbert entropy is effective in evaluating the fractal nature of the general high-dimensional complex data.

In summary, we introduce a new methodology, Hilbert entropy, for evaluating the complexity of high-dimensional physical systems. The application of the information entropies followed by the dimension reduction via the Hilbert curve enables an accurate specification of critical points in various phase transition phenomena. Furthermore, the Hilbert entropy, characterized by its power-law dependencies, showed a consistent linear relationship with the fractal and Euclidean dimensions in scale-invariant high-dimensional geometries. In particular, we have shown that the proposed method can be utilized for accurately assessing the fractal dimension of general systems, such as grayscale images. This method introduces a new approach to understanding and analyzing complex systems, potentially laying the groundwork for innovative strategies in analyzing and measuring complexity in high-dimensional physical systems.

**Methods**

LZ entropy

The Lempel-Ziv (LZ) complexity defines the complexity by assessing the pattern and structure of the given data sequence. Since the LZ complexity, $C_{LZ}$, is limited to its intrinsic property of being proportional to the size of the data, the normalized $C_{LZ}$, or LZ entropy ($S_{LZ}$), which is normalized by the maximum complexity of the given sequence, is proposed. The $C_{LZ}$ approaches the Kolmogorov complexity when the data size ($N$) is sufficiently large. The detailed calculation of $C_{LZ}$ can be found elsewhere[25]. Although $C_{LZ}$ is known for being suitable for measuring the complexity of the binary data, it can be extended and generalized to the $n$-digit data as follows:

$$S_{LZ} = \frac{C_{LZ} \log_n N}{N} \qquad (6)$$

Sample entropy

The sample entropy ($S_{samp}$) measures the complexity of the system based on the similarity between the sampled data with the same size. For the determination of the similarity, the Chebyshev distance is usually used. The given sequence ($u$) is divided into sub-sequences ($\mathbf{X}_m$) with the embedding dimension of $m$ defined as $\mathbf{X}_m(i) = \{u(i+k) | 0 \leq k \leq m-1\}$; if the distance between two sub-sequences is within the threshold distance $r$, they are assumed to be similar. Typically, $r$ is set to $r = 0.2\sigma$, where $\sigma$ is the standard deviation of the given sequence. The sample entropy $S_{samp}$ is defined as the ratio of the number of similar sub-sequences with respect to the $m$ as follows:

$$S_{\text{samp}} = -\log \frac{(N-m)\sum_{j=1,j\neq i}^{N-m} \#\{i|\mathbf{X}_{m+1}(i) \text{ s.t. } d[\mathbf{X}_{m+1}(i),\mathbf{X}_{m+1}(j)]<r\}}{(N-m)\sum_{j=1,j\neq i}^{N-m} \#\{i|\mathbf{X}_m(i) \text{ s.t. } d[\mathbf{X}_m(i),\mathbf{X}_m(j)]<r\}}, \quad (7)$$

where $d[\mathbf{X},\mathbf{X}^*]$ denotes the distance between two sub-sequences $\mathbf{X}$ and $\mathbf{X}^*$.

## Permutation entropy

The permutation entropy ($S_{\text{perm}}$) can be measured considering the probability of occurrence of the ordinal patterns. For this, the original signal is divided into partition $\mathbf{X}_m(i)$ with the size of the embedding dimension $m$. To calculate $S_{\text{perm}}$, the probability of a specific ordinal pattern occurring ($p(\pi)$) is defined as

$$p(\pi) = \frac{\#\{i|\mathbf{X}_m(i) \text{ s.t. type } \pi\}}{N-m+1}. \quad (8)$$

The permutation entropy is then measured by making ordinal patterns from the sampled data and subsequent implementation of Shannon's information entropy to the $p(\pi)$ such that

$$S_{\text{perm}} = -\sum p(\pi) \log p(\pi). \quad (9)$$

## Spin models

We used periodic boundary conditions to a $N^2$ square lattice ($N=2^7$) with the unit spacing for the Metropolis algorithm-based Monte Carlo (MC) simulation of 2D square lattice Ising and XY models. The maximum number of iterations was set to $10000N^2$. The energy of the system is measured for the temperature from 0 K to 10 K, with an interval of 0.1 K.

Ising model

To calculate the theoretical critical temperature on the 2D square lattice Ising model, we calculated the energy of spin states given by a Hamiltonian such that

$$\mathcal{H}(\sigma) = -J \sum_{\langle i,j \rangle} \sigma_i \sigma_j, \tag{10}$$

where $J$ denotes and interaction between adjacent sites $i$ and $j$, and $\sigma$ is a binary spin state such that $\sigma \in \{-1, +1\}$. For simplification, the interaction between neighboring spins, $J$, is set to 1. The external magnetic stimuli were neglected. The spin state of each spin is determined by the Boltzmann's probability distribution as follows:

$$P(\sigma) = \frac{e^{-\beta \Delta \mathcal{H}(\sigma)}}{Z}, \tag{11}$$

where $\beta = 1/k_B T$ is the inverse temperature and $Z$ is a partition function given by

$$Z = \sum_{\sigma} e^{\beta \mathcal{H}(\sigma)}. \tag{12}$$

The given system is initialized into a random spin configuration of -1 and +1 at the initial stage. For each iteration, the probability of flipping spins is governed by Boltzmann's probability distribution (equation (11)). The current spin state is replaced by the new spin state if the current state satisfies either $\Delta \mathcal{H}(\sigma) \leq 0$ or $p \leq P(\sigma)$ for $p \in [0,1]$.

The entropy $S$ of the system is then derived from the free energy $F$ with the free energy following Onsager's solution[28]:

$$S = -\frac{\partial F}{\partial T}$$
$$F = -\frac{1}{\beta}\left(\log 2 + \frac{1}{8\pi^2}\int_0^{2\pi}\int_0^{2\pi} \log\left[\cosh^2 2\beta J - \sinh 2\beta J\left(\cos\omega_1 + \cos\omega_2\right)\right]d\omega_1 d\omega_2\right). \quad (13)$$

The results are shown in Fig. 2b.

## XY model

The energy of the classical 2D XY model of unit interaction strength without an external magnetic field is given by

$$\mathcal{H}(\theta) = -J\sum_{\langle i,j \rangle} \cos(\theta_i - \theta_j), \quad (14)$$

where $\theta_i$ is a spin angle of $i$th spin accompanied by $j$ neighboring spins such that $\theta \in [-\pi, \pi]$ [38].

During the MC process, spin orientation suffers random variation of $\theta_i = \theta_i + 2\pi k$ for $k \in [-0.5, 0.5]$. The spin configuration is then determined by following Boltzmann's probability distribution (equation (11)) with the partition function[39]

$$Z = \int \prod_j e^{-\beta \mathcal{H}(\theta)} d\theta_j. \quad (15)$$

Since the analytic solution of the 2D XY model is unknown, rather than comparing the thermodynamic entropy, the critical point of the XY model can be analyzed by introducing the order parameter. Concerning the continuous spin of the XY model, the orientational correlation $g_\sigma(r)$ is expressed as follows:

$$g_\sigma(r) = \langle \Psi(\mathbf{r}_i) \cdot \Psi^*(\mathbf{r}_j) \rangle \quad (16)$$

where $r = |\mathbf{r}_i - \mathbf{r}_j|$ and $\Psi(\mathbf{r}_k) = e^{i\theta_k}$ for $k$th spin.

Following the relationship $g_\sigma(r) \sim r^{-\eta}$, the critical exponent $\eta$ can be evaluated from the slope of the envelope function against $r$ for each temperature $k_B T$ (Fig. 2d). The slope ($\eta$) abruptly increases at the transition temperature of 0.8972, which is close to the reported critical point of 0.8935[40].

## Chaos game-based fractal geometry

We generated IFS fractals in 2D/3D spaces based on the chaos game[41]. To make the size of the system uniform, the regular polygon/polyhedrons were inscribed in the unit circle/sphere. We generated polygons by adjusting the number of edges ($E \in \{3, 4, \cdots, 50\}$) or faces ($F \in \{4, 6, 8, 12, 20\}$). During the iteration process, the additional points are generated inside the initial polygon/polyhedron, following the iterative function:

$$\mathbf{x}_i = r\mathbf{x}_{i-1} + (1-r)\mathbf{b}_j, \tag{17}$$

where $\mathbf{x}_i$ is the coordinate of the point for $i$th iteration, $\mathbf{b}_j$ is the coordinate of the randomly chosen $j$th vertices where $j \in \{1, 2, \cdots, n\}$ for total number of vertices of $n$, and $r$ is a scale factor given by

$$r = \frac{1}{2\left(1 + \sum_{k=1}^{\lfloor n/4 \rfloor} \cos \frac{2\pi k}{n}\right)}, \tag{18}$$

which is applicable to 2D geometry[42].

For $n = 4$ 2D IFS, the theoretical fractal dimension is equal to 2 (i.e., simple square); therefore, we used Sierpinski carpet whose fractal dimension is $\frac{\log 8}{\log 3} = 1.8928$, for $n = 4$ 2D IFS.

## Acknowledgements

This work was supported by the Samsung Research Funding Center for Samsung Electronics under project number SRFC-MA2201-02.

## Author information

These authors contributed equally: Seong-Gyun Im and Taewoo Kang.

Authors and Affiliations

**School of Chemical Engineering, Sungkyunkwan University (SKKU), Republic of Korea**

Seong-Gyun Im, Taewoo Kang & S. Joon Kwon

**SKKU Institute of Energy Science & Technology (SIEST), SKKU, Republic of Korea**

S. Joon Kwon

**Department of Semiconductor Convergence Engineering, SKKU, Republic of Korea**

S. Joon Kwon

**Department of Advanced Energy Engineering, SKKU, Republic of Korea**

S. Joon Kwon

Contributions

S.-G.I. and T.K. performed the numerical simulations and analysis. S.-G.I., T.K., and S.J.K. wrote the manuscript. All authors were involved in editing and reviewing the paper.

Corresponding authors

Correspondence to S. Joon Kwon.

**Ethics declarations**

Competing interests

The authors declare no conflict of interest.

# Supplementary Information

# Hilbert entropy for measuring the complexity of high-dimensional systems

Seong-Gyun Im [1,†], Taewoo Kang [1,†], and S. Joon Kwon [1,2,3,4]*

[1]School of Chemical Engineering, Sungkyunkwan University (SKKU), Republic of Korea

[2]SKKU Institute of Energy Science & Technology (SIEST), SKKU, Republic of Korea

[3]Department of Semiconductor Convergence Engineering, SKKU, Republic of Korea

[4]Department of Advanced Energy Engineering, SKKU, Republic of Korea

* Corresponding author: sjoonkwon@skku.edu

[†] These authors contributed equally to this work.

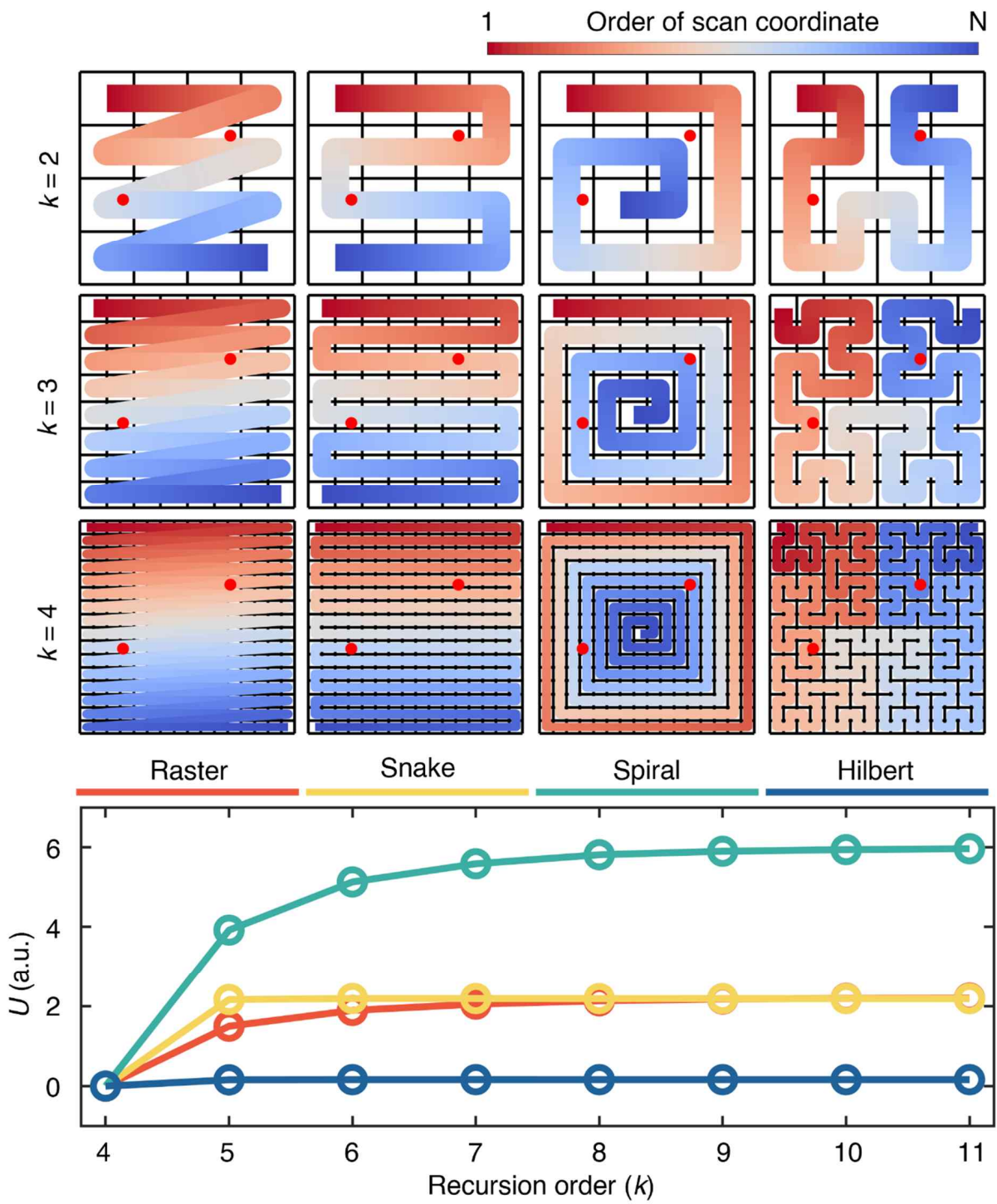

To test the performance of the scanning method based on the Hilbert curve, we compared the information uncertainty between two randomly selected points on the mapping function for the different scanning methods,

$$U = \text{std}\left[\left\|\frac{|q_{1,k} - q_{2,k}|}{2^{nk}} - \frac{|q_{1,\kappa} - q_{2,\kappa}|}{2^{n\kappa}}\right\|\right],$$

where $\kappa = \min\{k \in \mathbb{N} \mid k \leq M\}$, $q$ is the position of the points on the flattened data, and $M$ is the size of the lattice. Supplementary Fig. 1 shows that the Hilbert curve-based dimension reduction exhibits nearly zero uncertainty for any randomly selected points pair (the number of test pairs = $10^3$), while other methods produce fluctuation in uncertainties when the lattice cell size changes (i.e., fluctuation increases with increasing the recursion order $k$).

a Asymmetric  b Spherical  c 3D

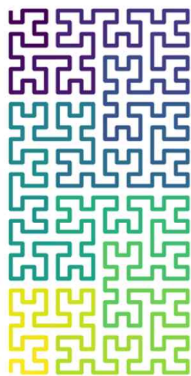 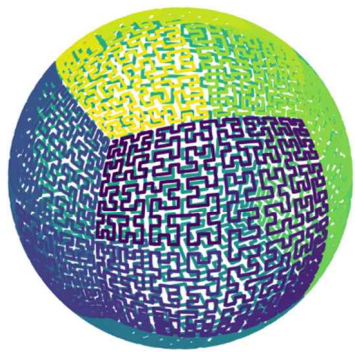 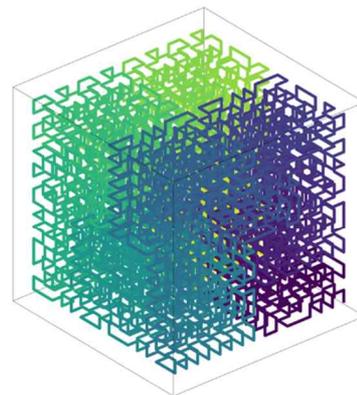

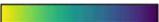

Order of scan coordinate

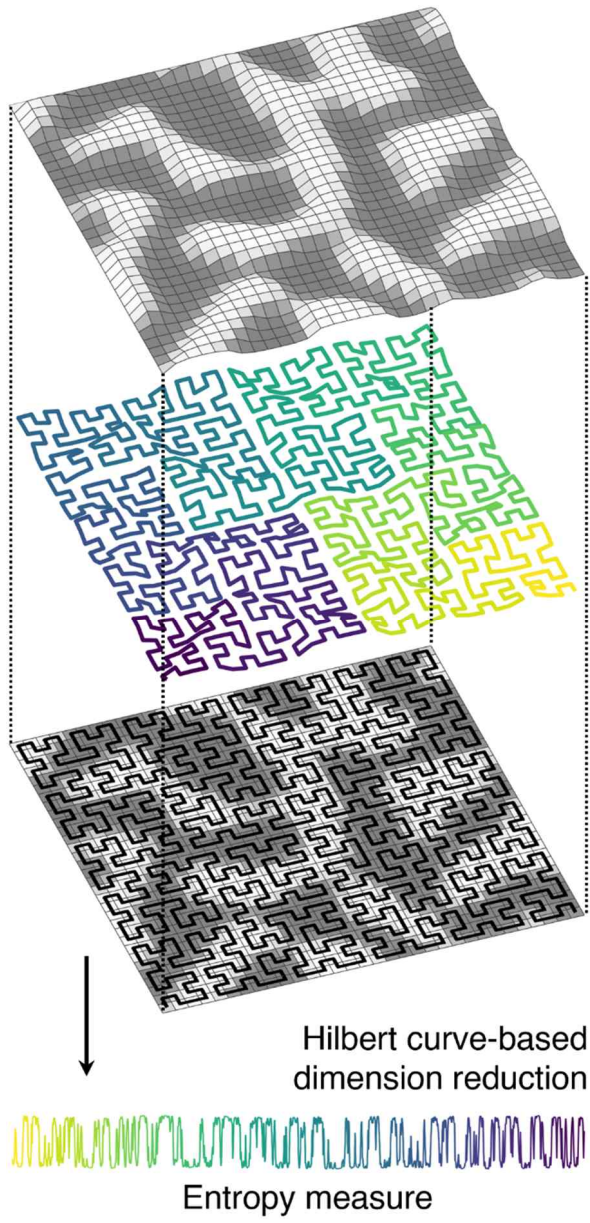

Hilbert curve-based
dimension reduction

Entropy measure

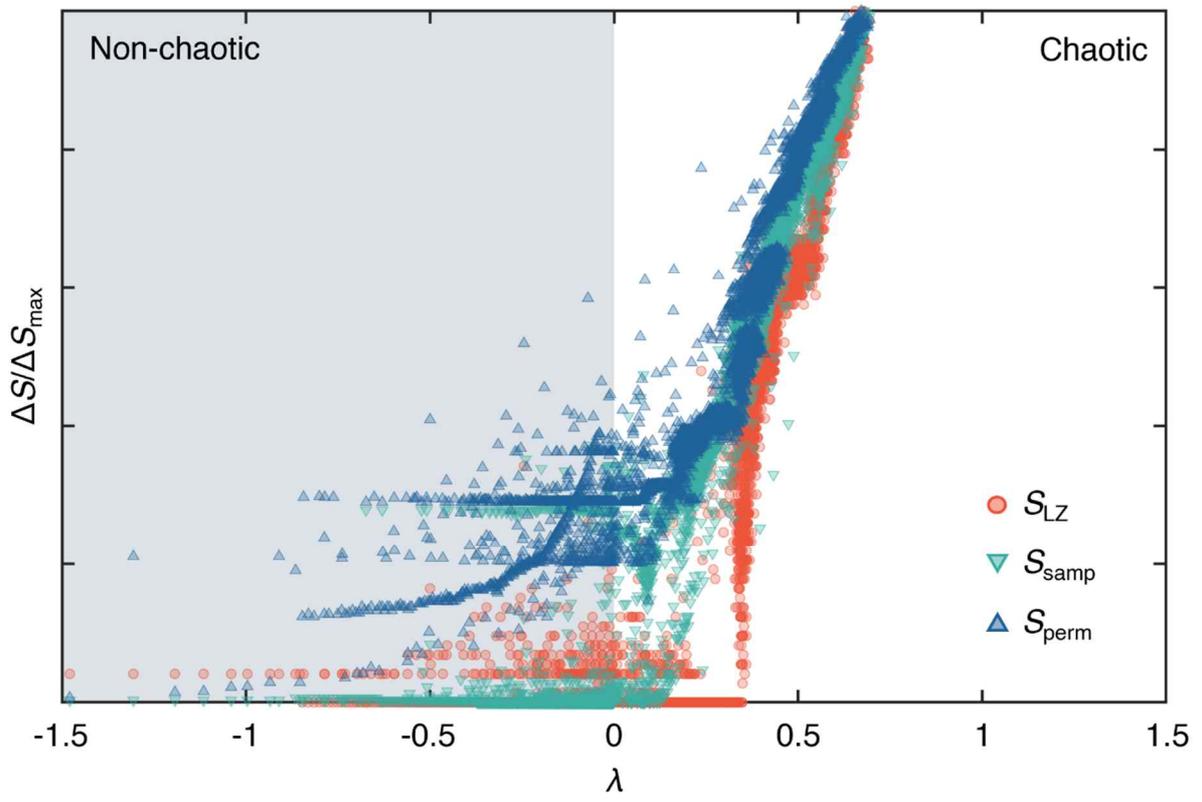

**a**

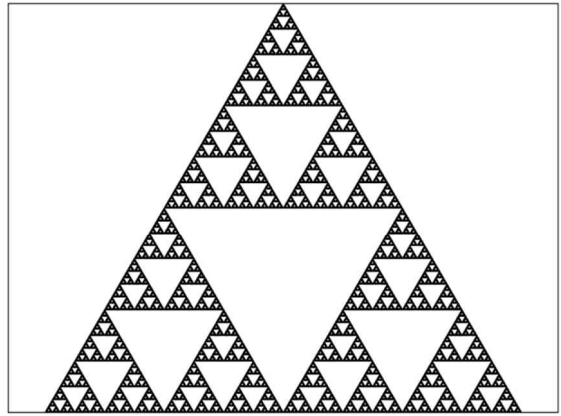

**b**

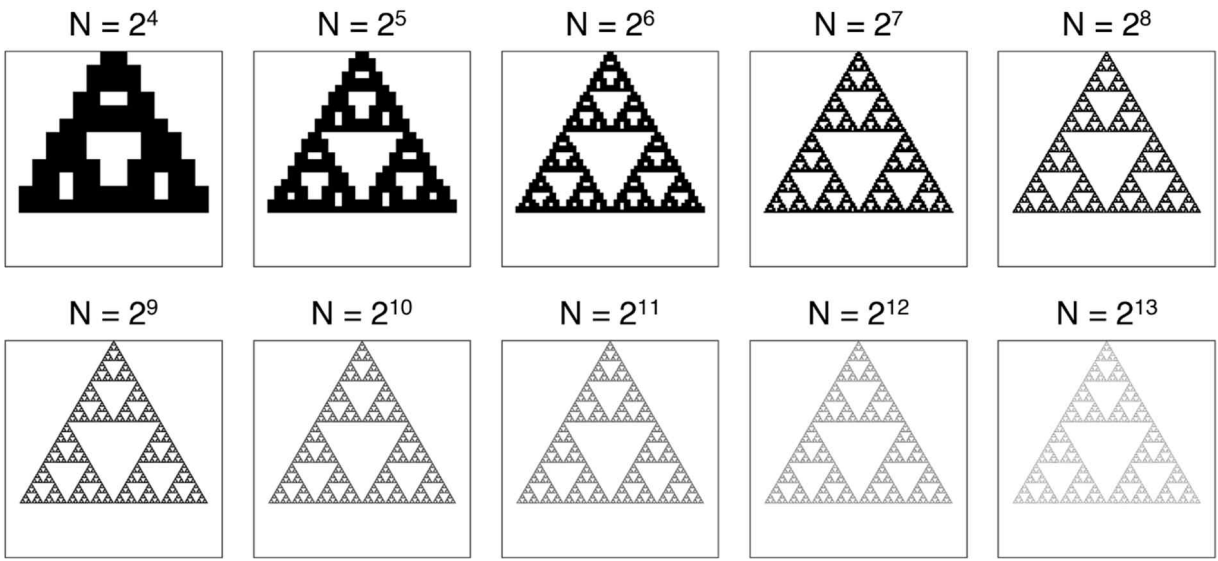

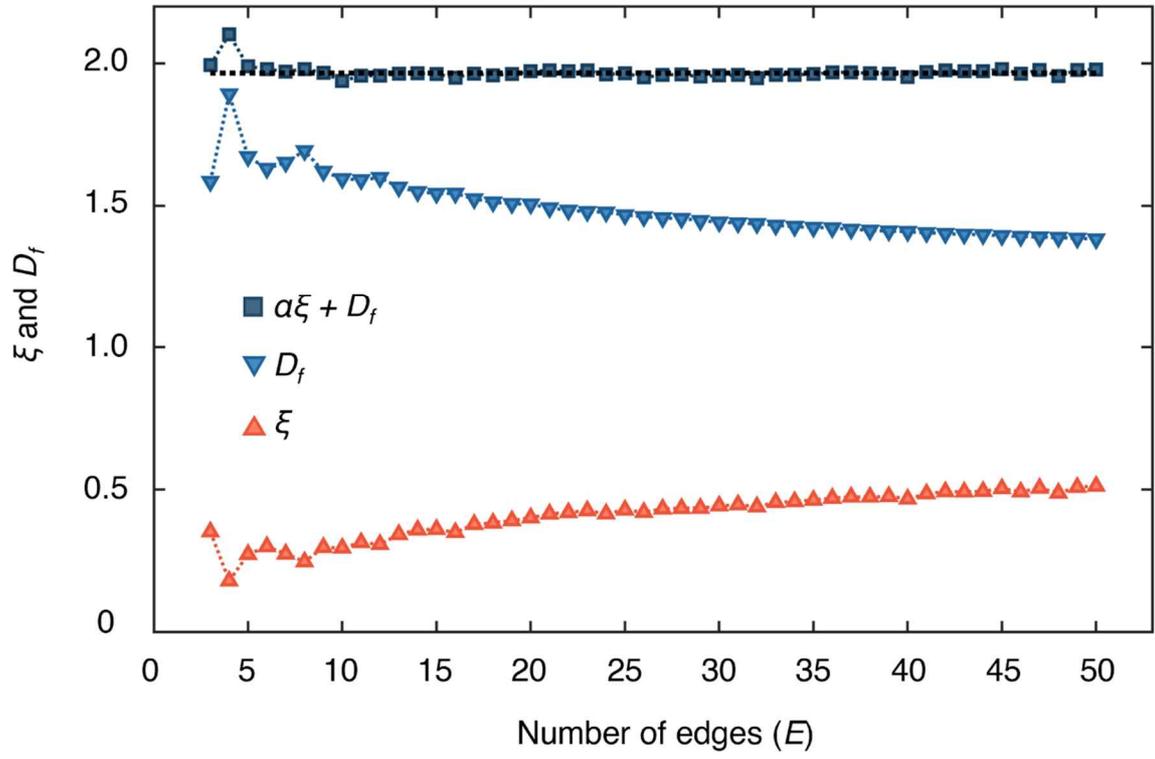

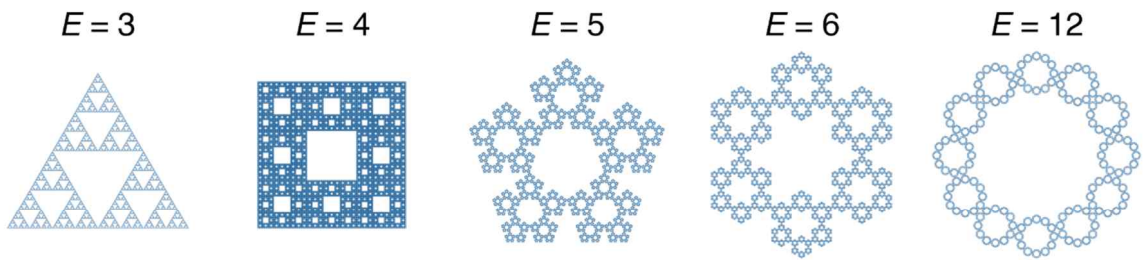

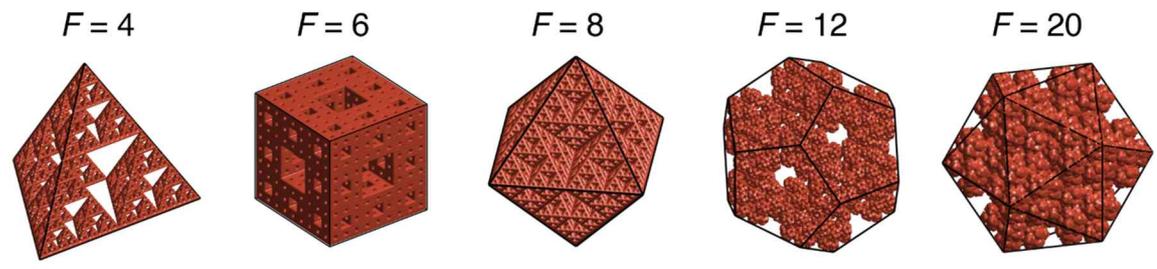

$\xi$ $D_f$

$E$

$D_f$    $D_f = \log(n)/\log(1/r)$   $n$   $r$

$\alpha$

$D_f = -\xi + D$    $D$

fractals with different number of edges or faces in 2D (**b**) and 3D (**c**). For 3D IFS, the number of faces ($F$) are adjusted while the number of edges is controlled in 2D IFS.